%% file: lenain_HEPRO_proc.tex
\documentclass[square]{ws-procs11x85}
\usepackage{multicol,float} 
\usepackage{psfrag}

\begin{document}

\include{aas_macros}

\title{SSC scenario for TeV emission from non-blazar AGNs}

\author{J.-P. LENAIN$^*$, C. BOISSON and H. SOL}

\address{LUTH, Observatoire de Paris, CNRS, Universit\'e Paris Diderot;\\
5 Place Jules Janssen, 92190 Meudon, France\\
$^*$E-mail: jean-philippe.lenain@obspm.fr}

\begin{abstract}
M\,87 is the first extragalactic source detected in the TeV $\gamma$-ray domain that is not a blazar, its large scale jet not being aligned to the line of sight. We present here a multi-blob synchrotron self-Compton model accounting explicitly for large viewing angles and moderate Lorentz factors as inferred from magnetohydrodynamic simulations of jet formation, motivated by the detection of M\,87 at very high energies (VHE; $E > 100$\,GeV). Predictions are presented for the very high-energy emission of active galactic nuclei with extended optical or X-ray jet, which could be misaligned blazars but still show some moderate beaming. We include predictions for 3C\,273, Cen\,A and PKS\,0521$-$36.
\end{abstract}

\keywords{Galaxies: active; Galaxies: individual (M\,87, 3C\,273, Cen\,A, PKS\,0521$-$36); Gamma rays: theory; Radiation mechanisms: non-thermal.}

\bodymatter

\begin{multicols}{2}

\section{H.E.S.S.}\label{jpl-sec-hess}

The H.E.S.S. (High Energy Stereoscopic System) collaboration operates an array of four 13\,m imaging atmospheric \v{C}erenkov telescopes. The cameras measure $\gamma$-rays above a threshold of about 100\,GeV up to 10\,TeV, depending on the zenith angle of the source at the observing time, by imaging the \v{C}erenkov light induced by an air shower developing when a VHE photon or particle enters the atmosphere. The stereoscopy technique allows a precise reconstruction of the arrival position and energy of the incident VHE photon, as well as a good discrimination between hadrons and photons (see~\cite{2005A&A...430..865A} for more details). The recent observations of M\,87 by H.E.S.S.\cite{2006Sci...314.1424A} motivated us to develop the multi-blob radiative model presented below.

\section{The multi-blob model}\label{jpl-sec-multiblob}

The classic synchrotron self-Compton (SSC) models have difficulties accounting for the VHE emission of sources with misaligned jets, unless very high, unphysical values for the Doppler factor are assumed. We present here a model for the jet emission in active galactic nuclei (AGNs) based on the model from~\cite{2001A&A...367..809K,2003A&A...410..101K}, and accounting explicitly for low Lorentz factors with an emitting zone close to the central supermassive black hole\cite{2008A&A...478..111L}.

According to the results of general relativistic magnetohydrodynamic simulations for the bulk Lorentz factor of jets, the strength of the magnetic field at the Alfv\'en surface, and the opening angle of the jet\cite{2006MNRAS.368.1561M}, we can put constraints on these macrophysic parameters in our model. The emitting zone is represented by a cap in the broadened formation zone at the base of the jet, located just downstream of the Alfv\'en surface to leave enough time for the acceleration processes to develop. This cap is filled with several blobs of plasma moving in slightly different directions, inducing a differential Doppler effect between these blobs. The radiative transfer is then computed for each blob in its own source frame. For more details, see~\cite{2008A&A...478..111L}.

\begin{figure}[H]
  \psfrag{theta}{\footnotesize $\theta$}
  \psfrag{jet axis}{jet axis}
  \psfrag{rb}{\footnotesize $r_b$}
  \psfrag{Gamma_b}{\footnotesize $\Gamma_b$}

  \centerline{\psfig{file=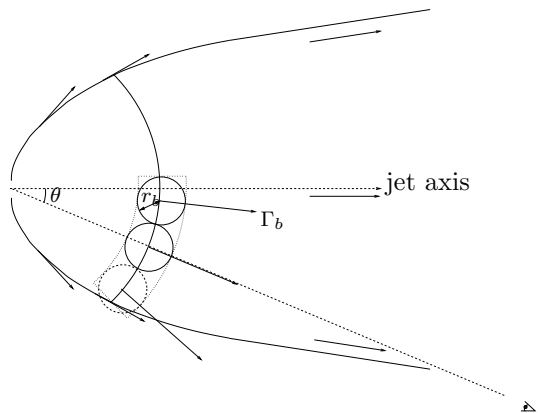,width=7cm}}
  \caption{Geometrical scheme of the multi-blob model.}\label{jpl-fig1}
\end{figure}

\section{M\,87}\label{jpl-sec-M87}

\begin{figure}[H]
  \centerline{\psfig{file=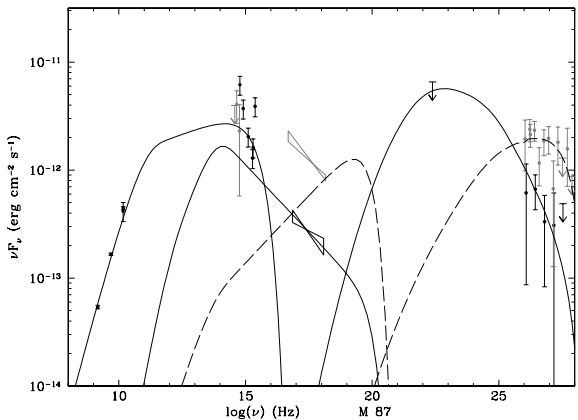,width=6cm,angle=-90}}
  \caption{Spectral energy distribution of M\,87 with the multi-blob SSC model. The peak in radio represents the synchrotron emission from an extended jet. The bumps in optical/X-ray and in $\gamma$-rays are the sum of the contributions of the different blobs through the SSC process ({\it solid and dashed lines}).}\label{jpl-fig2}
\end{figure}

We apply this model to the case of M\,87 ($z=0.00436$), which is the first extragalactic object observed in the VHE range with a misaligned jet (see Fig.~\ref{jpl-fig2}). The solid line shows the best solution for the low state of activity represented by the {\it Chandra\/} data\cite{2003ApJ...599L..65P} of 2000 along with the H.E.S.S. data\cite{2006Sci...314.1424A} of 2004. The corresponding parameters are given in~\tref{jpl-tbl1}, where $\Gamma_b$ is the bulk Lorentz factor common to each blob, $\theta$ is the viewing angle of the jet, $r_g = G M_\mathrm{BH}/c^2$ is the scale length with $M_\mathrm{BH}$ the mass of the central black hole, $B$ is the magnetic field, $r_b$ stands for the radius of the blobs, and the other parameters describe the population of electrons assumed to be a broken power law, with $n_1$ and $n_2$ the spectral indices before and after the break.

The dashed lines represent a solution for the higher state of activity observed by H.E.S.S. in 2005. Unfortunately, no simultaneous X-ray data are available to further constrain the model. However, this solution illustrates the ability of the multi-blob model to generate spectra sufficiently hard in the VHE range to reproduce the most recent H.E.S.S. observations. Interestingly, for both solutions we obtain a characteristic scale length of the order of the gravitational radius of the central supermassive black hole ($r_g \sim 4.5 \times 10^{14}$\,cm for $M_\mathrm{BH} \sim 3 \times 10^9 M_\odot$ \cite{1997ApJ...489..579M}), along with moderate values for the bulk Lorentz factor. Such small features can develop beyond the Alfv{\'e}n surface due to turbulence or reconfined shocks\cite{2006MNRAS.368.1561M} (see~\cite{2008A&A...478..111L} for a more detailed discussion).

\begin{table*}
  \tbl{Parameters used in Figs.~\ref{jpl-fig2}, \ref{jpl-fig3}, \ref{jpl-fig4}, and \ref{jpl-fig5}}
      {\begin{tabular}{@{}ccccccc@{}}
          \toprule
          Object & M\,87 & M\,87 & 3C\,273 & Cen\,A & Cen\,A & PKS\,0521$-$36\\
          Figure & \fref{jpl-fig2} ({\it solid\/}) & \fref{jpl-fig2} ({\it dashed\/}) & \fref{jpl-fig3} & \fref{jpl-fig4} ({\it dashed\/}) & \fref{jpl-fig4} ({\it solid\/}) & \fref{jpl-fig5}\\\colrule
          $\Gamma_b$ & 10.0 & 10.0 & 7.4 & 8.14 & 20.0 & 1.5\\
          $\theta$ & 15$^\circ$ & 15$^\circ$ & 15$^\circ$ & 25$^\circ$ & 25$^\circ$ & 25$^\circ$\\
          $B$ [G] & 0.01 & 0.01 & 3.0 & 2.0 & 10.0 & 1.0\\
          $r_b$ [cm] & $2.8 \times 10^{14}$ & $8.0 \times 10^{13}$ & $2.0 \times 10^{15}$ & $1.0 \times 10^{14}$ & $8.0 \times 10^{13}$ & $9.0 \times 10^{14}$\\
          $K_1$ [cm$^{-3}$] & $1.8 \times 10^4$ & $2.2 \times 10^4$ & $1.8 \times 10^6$ & $9.0 \times 10^7$ & $4.0 \times 10^4$ & $3.0 \times 10^6$\\
          $n_1$ & 1.5 & 1.5 & 2.0 & 2.0 & 2.0 & 2.0\\
          $n_2$ & 3.5 & 2.5 & 4.1 & 3.0 & 3.5 & 2.5\\
          $\gamma_\mathrm{min}$ & $10^3$ & $10^3$ & $1$ & $3.0 \times 10^2$ & $10^3$ & $10^3$\\
          $\gamma_\mathrm{br}$ & $10^4$ & $10^4$ & $1.6 \times 10^3$ & $5.0 \times 10^2$ & $3.5 \times 10^5$ & $5.0 \times 10^4$\\
          $\gamma_c$ & $10^7$ & $10^7$ & $10^6$ & $4.0 \times 10^3$ & $6.0 \times 10^6$ & $10^6$\\\botrule
      \end{tabular}}
      \label{jpl-tbl1}
\end{table*}

\section{Predictions of VHE flux for three AGNs}\label{jpl-sec-predictions}

We also apply the multi-blob model to three AGNs with extended optical/X-ray jet, which could be misaligned blazars. A more detailed description can be found in~\cite{2008A&A...478..111L}.

\begin{figure}[H]
  \centerline{\psfig{file=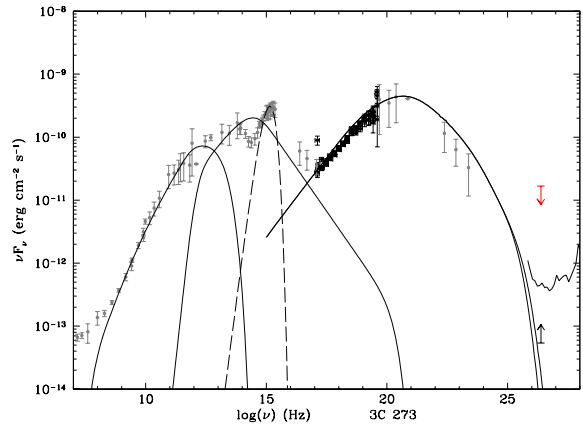,width=6cm,angle=-90}}
  \caption{Spectral energy distribution of 3C\,273 with anticipated VHE flux. The expected sensitivity of the next generation CTA project of $\sim 0.1\%$\,Crab flux at 1\,TeV for 50\,h of observation is shown as a lower limit. This source should be detectable with CTA.}\label{jpl-fig3}
\end{figure}

\subsection{3C\,273}

3C\,273 ($z=0.158$) is a well-known quasar, whose jet viewing angle is at most $15^\circ$\cite{1985ApJ...289..109U}. In~\fref{jpl-fig3}, we present the spectral energy distribution (SED) with the anticipated VHE flux as computed with the multi-blob model. The radio is modeled by synchrotron emission from the extended jet, thought to harbor a different population of electrons than the VHE-emitting blobs (solid line). The dashed line in the UV represents a simple blackbody model illustrating the contribution from the accretion disk. In~\fref{jpl-fig3}, the solid curve in the VHE range represents the H.E.S.S. sensitivity at $5\sigma$ for 50\,h of observing a source at a mean zenith angle of $30^\circ$. It is thus obvious that a strong detection of 3C\,273 in the VHE range by the current \v{C}erenkov facilities would be difficult to interpret within our multi-blob scenario. An alternative would be to assume the presence of different electron energy distributions among the blobs, resulting in a tail at high energy for the inverse Compton bump, or the presence of an external inverse Compton component, which is then expected to be only weakly variable. Further observations with HESS~2 and {\it GLAST} are crucial to distinguish between the different scenarii.

\subsection{Cen\,A}

Cen\,A is a nearby very well-studied radiogalaxy ($z=0.0018$). The value of the viewing angle of its jet depends on the scale it is looked at. On the parsec-scale jet, \cite{1998AJ....115..960T} found $\theta \sim 50^\circ$--$80^\circ$, whereas \cite{2003ApJ...593..169H} found $\theta \sim 15^\circ$ on the 100\,pc scale jet. We choose to take an intermediate value of $\theta \sim 25^\circ$.

\begin{figure}[H]
  \centerline{\psfig{file=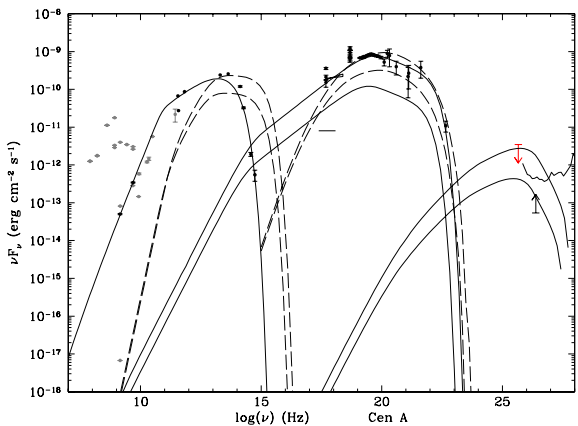,width=6cm,angle=-90}}
  \caption{Spectral energy distribution of Cen\,A with anticipated VHE flux within the multi-blob SSC model.}\label{jpl-fig4}
\end{figure}

The nature of the soft $\gamma$-ray emission is still an unsolved issue. Assuming an inverse Compton origin, we obtain the solution shown in dashed lines in~\fref{jpl-fig4}. In that case, the SSC emission would cut off before the VHE regime. A detection of Cen\,A at VHE would then favor the presence of an external inverse Compton component, e.g. from the starlight photons of the host galaxy.

However, if the soft $\gamma$-rays are due to synchrotron emission, the solution in solid lines shows that we can expect a marginal detection of the core of Cen\,A within 50\,h of observation by H.E.S.S.

\subsection{PKS\,0521$-$36}

PKS\,0521$-$36 ($z=0.0553$) is an object ``oscillating'' between a Seyfert-like and a BL\,Lac-like state\cite{1981A&A...103L...1U}, thus difficult to model within a purely non-thermal scenario, especially since the available data are not contemporaneous.

\begin{figure}[H]
  \centerline{\psfig{file=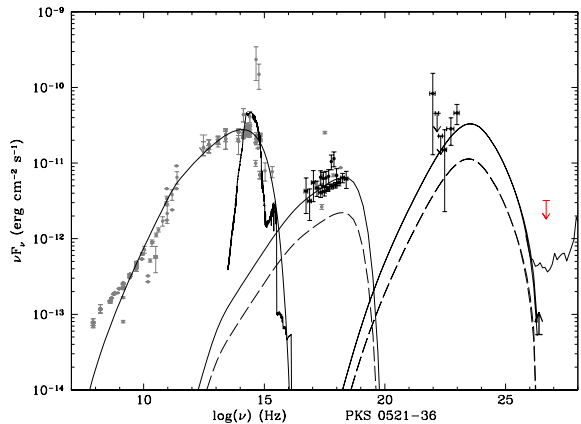,width=6cm,angle=-90}}
  \caption{Spectral energy distribution of PKS\,0521$-$36 with the multi-blob scenario.}\label{jpl-fig5}
\end{figure}

In~\fref{jpl-fig5}, the black line in the optical shows a simple modeling for the thermal emission of the host galaxy. The solid and dashed lines represents two extreme geometrical situations within the multi-blob scenario. In one case ({\it solid line}), the line of sight is aligned with the velocity vector of a blob. In the other one ({\it dashed line}), it passes exactly through the gap between three blobs.

If the X-rays were due to inverse Compton scattering, then most likely on photons from the radio/optical extended jet, this contribution would not be highly variable. Since the X-rays are strongly varying in this source, we assume that they come from synchrotron emission. In this case, PKS\,0521$-$36 should be marginally detectable by H.E.S.S., and easily detectable by H.E.S.S.~2 and CTA.

\section{Conclusions}

We developed an SSC model to interpret the variable VHE emission observed in M\,87 and to predict the VHE emission for three AGNs showing an extended jet. This model accounts in a simple way for differential Doppler-boosting by modeling the radiative transfer of several blobs in the broadened formation zone of the jet, and provides the possibility to extend classical leptonic SSC models of TeV blazars to different types of AGNs with misaligned jet, leading to moderate values of the bulk Lorentz factor of the jet.

When applied to very different radio-loud objects, the multi-blob model deduces very similar properties for the size of the VHE emitting zone and the strength of the magnetic field. The inferred bulk Lorentz factor $\Gamma_b$ can remain below a value of 10 for misaligned objects, thus reconciling SSC models with (GR)MHD models for the formation of jets. This model thus seems to fit well within a unified picture of AGNs\cite{1995PASP..107..803U}. Furthermore, our scenario brings a potential solution to the long-standing paradox of the apparent absence of high superluminal motion at the base of radio jets of TeV BL\,Lacs (see~\cite{2008A&A...478..111L} for a more detailed discussion on this issue). In our model, some X- and $\gamma$-rays emitting plasma blobs are moving very close to the line of sight and dominate the flux emission, allowing Cen\,A and PKS\,0521$-$36, for instance, to be potentially detected in VHE, while harboring misaligned extended jets.

\section*{Acknowledgments}

J.-P.~L. would like to thank Dr.~A.~Djannati-Ata{\"i}, Dr.~S.~Pita, and Dr.~A.~Zech for useful discussions.

This research made use of the NASA/IPAC Extragalactic Database (NED), which is operated by the Jet Propulsion Laboratory, California Institute of Technology, under contract with the National Aeronautics and Space Administration.

\bibliographystyle{ws-procs11x85}
\bibliography{lenain_HEPRO_proc}
\end{multicols}
\end{document}

%% file: aas_macros.tex
%
%
%


\def\jnl@style{\it}
\def\aaref@jnl#1{{\jnl@style#1}}

\def\aaref@jnl#1{{\jnl@style#1}}

\def\aj{\aaref@jnl{AJ}}                   
\def\araa{\aaref@jnl{ARA\&A}}             
\def\apj{\aaref@jnl{ApJ}}                 
\def\apjl{\aaref@jnl{ApJ}}                
\def\apjs{\aaref@jnl{ApJS}}               
\def\ao{\aaref@jnl{Appl.~Opt.}}           
\def\apss{\aaref@jnl{Ap\&SS}}             
\def\aap{\aaref@jnl{A\&A}}                
\def\aapr{\aaref@jnl{A\&A~Rev.}}          
\def\aaps{\aaref@jnl{A\&AS}}              
\def\azh{\aaref@jnl{AZh}}                 
\def\baas{\aaref@jnl{BAAS}}               
\def\jrasc{\aaref@jnl{JRASC}}             
\def\memras{\aaref@jnl{MmRAS}}            
\def\mnras{\aaref@jnl{MNRAS}}             
\def\pra{\aaref@jnl{Phys.~Rev.~A}}        
\def\prb{\aaref@jnl{Phys.~Rev.~B}}        
\def\prc{\aaref@jnl{Phys.~Rev.~C}}        
\def\prd{\aaref@jnl{Phys.~Rev.~D}}        
\def\pre{\aaref@jnl{Phys.~Rev.~E}}        
\def\prl{\aaref@jnl{Phys.~Rev.~Lett.}}    
\def\pasp{\aaref@jnl{PASP}}               
\def\pasj{\aaref@jnl{PASJ}}               
\def\qjras{\aaref@jnl{QJRAS}}             
\def\skytel{\aaref@jnl{S\&T}}             
\def\solphys{\aaref@jnl{Sol.~Phys.}}      
\def\sovast{\aaref@jnl{Soviet~Ast.}}      
\def\ssr{\aaref@jnl{Space~Sci.~Rev.}}     
\def\zap{\aaref@jnl{ZAp}}                 
\def\nat{\aaref@jnl{Nature}}              
\def\iaucirc{\aaref@jnl{IAU~Circ.}}       
\def\aplett{\aaref@jnl{Astrophys.~Lett.}} 
\def\apspr{\aaref@jnl{Astrophys.~Space~Phys.~Res.}}
\def\bain{\aaref@jnl{Bull.~Astron.~Inst.~Netherlands}} 
\def\fcp{\aaref@jnl{Fund.~Cosmic~Phys.}}  
\def\gca{\aaref@jnl{Geochim.~Cosmochim.~Acta}}   
\def\grl{\aaref@jnl{Geophys.~Res.~Lett.}} 
\def\jcp{\aaref@jnl{J.~Chem.~Phys.}}      
\def\jgr{\aaref@jnl{J.~Geophys.~Res.}}    
\def\jqsrt{\aaref@jnl{J.~Quant.~Spec.~Radiat.~Transf.}}
\def\memsai{\aaref@jnl{Mem.~Soc.~Astron.~Italiana}}
\def\nphysa{\aaref@jnl{Nucl.~Phys.~A}}   
\def\physrep{\aaref@jnl{Phys.~Rep.}}   
\def\physscr{\aaref@jnl{Phys.~Scr}}   
\def\planss{\aaref@jnl{Planet.~Space~Sci.}}   
\def\procspie{\aaref@jnl{Proc.~SPIE}}   

\let\astap=\aap
\let\apjlett=\apjl
\let\apjsupp=\apjs
\let\applopt=\ao